\documentclass[noeprint,bibnotes,amsmath,amssymb,superscriptaddress,aps,pra,reprint]{revtex4-1}
\usepackage{graphicx}
\usepackage{dcolumn}
\usepackage{amsmath}
\usepackage{amssymb}
\usepackage{latexsym}
\usepackage{epsfig}
\usepackage{amsbsy}
\usepackage{array}
\usepackage{setspace}
\usepackage{bm}
\usepackage{subfigure}
\usepackage{eqnarray}
\usepackage{hyperref}
\usepackage{color}
\hypersetup{colorlinks=true,linkcolor=red,urlcolor=blue,citecolor=blue
,anchorcolor=green}

\makeatletter

\newcommand{\Rmnum}[1]{\expandafter\@slowromancap\romannumeral #1@}
\makeatother
\begin{document}
\title{ Investigation of a non-Hermitian edge burst with time-dependent perturbation theory}
\author{Pengyu Wen}
\email{These authors contribute equally to this work.}
\affiliation{Department of Physics, Tsinghua University, Beijing 100084,China}
\author{Jinghui Pi}
\email{First and second author contribute equally to this work}
\email{pijh14@gmail.com}
\affiliation{Department of Physics, Tsinghua University, Beijing 100084,China}
\author{Gui-Lu Long}
\email{gllong@tsinghua.edu.cn}
\affiliation{Department of Physics, Tsinghua University, Beijing 100084,China}
\affiliation{Frontier Science Center for Quantum Information, Beijing 100084,China}
\affiliation{Beijing Academy of Quantum Information Sciences, Beijing 100193, China}
\date{\today}


\begin{abstract}
Edge burst is a phenomenon in non-Hermitian quantum dynamics discovered by a 
recent numerical study [\href{https://link.aps.org/doi/10.1103/PhysRevLett.128.120401}{W.-T. Xue, {\it et al}, Phys.\;Rev.\;Lett\;{\bf 2}, 128.120401(2022)}]. 
It finds that a large proportion of particle loss occurs at the system
boundary in a class of non-Hermitian quantum walk. In this paper, we investigate the evolution 
of real-space wave functions for this lattice system. We find the wave function of the edge site is distinct from the bulk sites. 
Using time-dependent perturbation theory, we derive the analytical expression of the real-space wave functions 
and find that the different evolution behaviors between the edge and bulk sites are due to their different nearest-neighbor site configurations.
 We also find the edge wave function primarily results from the transition of the two nearest-neighbor non-decay sites. Besides, the numerical diagonalization shows the edge wave function
 is mainly propagated by a group of eigen-modes with a
relatively large imaginary part. Our work provides an analytical method for studying non-Hermitian quantum dynamical problems.
\end{abstract}

\maketitle


\section{\label{sec:level1}INTRODUCTION}
The Hermiticity of the Hamiltonian is a fundamental requirement for a closed system in quantum physics \cite{shankar2012principles}. 
In many situations, however, one is only interested in a limited subspace of the whole system, 
and it can be encapsulated in an effective non-Hermitian Hamiltonian \cite{el2018non,ashida2020non}. 
Such systems include but are not limited to optical systems with gain and loss \cite{PhysRevLett.100.103904,PhysRevLett.101.080402,PhysRevLett.103.093902,
PhysRevLett.102.065703,PhysRevLett.103.123601,PhysRevLett.106.093902,regensburger2012parity,PhysRevLett.108.024101,PhysRevLett.108.173901,
doi:10.1126/science.1258479,doi:10.1126/science.1258480}, open systems with dissipation
\cite{PhysRevLett.70.2273,Rotter_2009,PhysRevLett.104.153601,PhysRevLett.106.213901,PhysRevX.4.041001,PhysRevLett.115.200402,PhysRevLett.123.123601,
li2019observation},
 and electron systems with finite-lifetime quasiparticles \cite{PhysRevB.97.041203,PhysRevLett.121.026403,PhysRevB.98.035141,
 doi:10.1126/science.aap9859,PhysRevB.99.201107,PhysRevB.100.115124}. 
Another important class of non-Hermitian systems is 
provided by lattices, where the role of topology has attracted tremendous interest \cite{PhysRevLett.116.133903,PhysRevLett.118.040401,
PhysRevX.8.031079,PhysRevLett.120.146402,PhysRevLett.121.026808,PhysRevLett.121.136802,PhysRevLett.121.086803,PhysRevB.97.121401,
PhysRevB.99.201103,PhysRevLett.123.016805,PhysRevResearch.1.023013,
PhysRevB.99.235112,PhysRevB.99.125103,PhysRevB.99.081302,PhysRevLett.123.246801,
PhysRevLett.123.170401,PhysRevLett.123.066404,PhysRevB.100.045141,PhysRevB.99.081103,PhysRevLett.122.076801,Ghatak_2019,PhysRevX.9.041015,
PhysRevA.99.052118,PhysRevLett.124.056802,Longhi:19,PhysRevB.99.245116,PhysRevB.100.165430,PhysRevLett.123.097701,
PhysRevLett.123.206404,PhysRevLett.122.237601,PhysRevLett.124.086801,PhysRevLett.125.186802,PhysRevB.102.205118,PhysRevB.101.045415,PhysRevB.101.195147,PhysRevLett.124.066602,
PhysRevLett.125.126402,PhysRevLett.125.226402,PhysRevLett.125.260601,lee2020ultrafast,li2020critical,PhysRevB.103.045420,PhysRevB.102.241202,
PhysRevB.102.201103,PhysRevResearch.2.043167,PhysRevResearch.2.043046,PhysRevLett.125.206402,xiao2020non,ghatak2020observation,
helbig2020generalized,PhysRevResearch.2.023265,doi:10.1126/science.aaz8727,PhysRevApplied.14.064076,FoaTorres_2020,RevModPhys.93.015005,Wang_2021,PhysRevLett.126.010401,
PhysRevB.103.165123,PhysRevB.103.085428,PhysRevLett.126.216407,PhysRevB.103.L140201,PhysRevLett.126.176601,PhysRevLett.126.216405,PhysRevLett.127.116801,PhysRevLett.127.066401,PhysRevLett.127.070402,doi:10.1126/science.abf6568,
PhysRevB.103.L241408,PhysRevB.103.205205,Lin:21,Longhi:21,PhysRevB.104.165117,PhysRevB.104.125109,wang2021topological,
PhysRevLett.127.270602,zhang2022universal,PhysRevB.105.045422,weidemann2022topological,PhysRevLett.129.113601,PhysRevLett.128.157601,
PhysRevLett.129.180401,PhysRevLett.128.223903,PhysRevB.106.035425,mittal2021persistence,wu2022complex,sigwarth2022time,xu2022exact}. 
A unique feature of non-Hermitian lattices is the non-Hermitian skin effect (NHSE), namely 
the localization of an enormous number of bulk-band eigenstates at the edges under open boundary conditions 
\cite{PhysRevLett.121.086803,PhysRevB.97.121401,PhysRevB.99.201103,PhysRevLett.123.016805,PhysRevResearch.1.023013,PhysRevLett.124.086801,
PhysRevLett.125.186802,PhysRevB.102.205118,PhysRevResearch.2.043167,PhysRevLett.126.176601,PhysRevLett.127.066401,PhysRevB.103.L140201,
PhysRevLett.127.116801,PhysRevLett.126.216405}. A significant consequence of the NHSE is the breakdown of the conventional
bulk-boundary correspondence, which can be recovered by using the localized skin modes replacing the extended Bloch waves of Hermitian
lattices \cite{PhysRevLett.121.026808,PhysRevLett.121.136802,PhysRevLett.121.086803,PhysRevLett.123.246801,helbig2020generalized,xiao2020non,ghatak2020observation,RevModPhys.93.015005}. 
It indicates that the boundary is even more important in non-Hermitian physics compared to their Hermitian counterparts.  

Recently, a novel boundary-induced dynamical phenomenon named "edge burst" is reported in Ref.~\cite{PhysRevLett.128.120401} and an experimental verification is reported in Ref.~\cite{Xiao2023Observation}. 
When a quantum particle (called "quantum walker") walks freely in a class of lossy lattices, 
it is intuitively expected that the decay probability is dominated by the lossy sites near the initial location of the walker. 
Surprisingly, the numerical study finds that there is an unexpected remarkable loss probability peak at the edge, with an almost invisible
loss probability tail in the bulk \cite{PhysRevLett.128.120401}. The appearance of such an edge peak has also been reported in an earlier work 
with an incorrect explanation \cite{Wang_2021}, which is based on topology. 
The edge burst was regarded as an interplay of the NSHE and the imaginary gap closing in 
Ref.~\cite{PhysRevLett.128.120401}, where the authors give a criterion based on the property of the Hamiltonian.
Although previous work has investigated the edge burst phenomenon, however,
 the relation between the real space dynamical behavior of the system and the formation of this burst edge lossy peak is still unconscious.

 In this work, we investigate the real-space dynamical behavior of quantum particles in a lossy lattice. 
 Numerical simulations show the edge burst phenomenon is closely related to the distinct dynamical behaviors of wave functions between the edge and bulk sites. To further understand how the walker propagates in the lattice, we introduce 
 time-dependent perturbation theory in non-Hermitian systems and evaluate the 
 analytical expression of the real-space quantum walk wave functions. Due to the NHSE,
 the walker mainly hops nonreciprocally along the non-decay chain. 
 The analysis of real-space wave functions shows the different evolution features between the edge
 and bulk can be attributed to their nearest-neighbor site configurations, which limits the
 possible path the walker can travel from the initial state. Besides, we find that 
 the main contribution to the edge wave function originates 
 from the interference transition of the two nearest-neighbor non-decay sites. 
 Furthermore, we discuss the relation between the evolution of the system and 
 its open (periodic) boundary eigen-modes  by numerical diagonalization. The result shows the walker is mainly propagated by a group of eigen-modes 
 which have a large imaginary part when it arrives at the edge. Our work gives an 
 explicit illustration of the edge burst phenomenon in real space and provides an alternative method to investigate non-Hermitian dynamical problems.

This paper is organized as follows. We introduce the non-Hermitian quantum walk model and describe the 
edge burst phenomenon with numerical simulations in Sec.~\ref{sec:level2}. A sketch of the time-dependent perturbation theory for a non-Hermitian 
Hamiltonian is given in Sec.~\ref{sec:level3}. We apply the theory in a concrete quantum walk model and solve the evolution equation 
analytically in Sec.~\ref{sec:level4}. By the analysis of wave functions, we elucidate the propagation process of the walker and 
the formation of edge burst in real space directly. In Sec.~\ref{sec:level5}, we discuss the relation between the evolution of the edge wave function and the eigenstates of the system. 
Finally, a summary and discussion are  given in Sec.~\ref{sec:level6}.

\section{\label{sec:level2} Model and non-Hermitian Edge Burst}

\begin{figure}[tbp]
     \centering
     \includegraphics[scale=0.28]{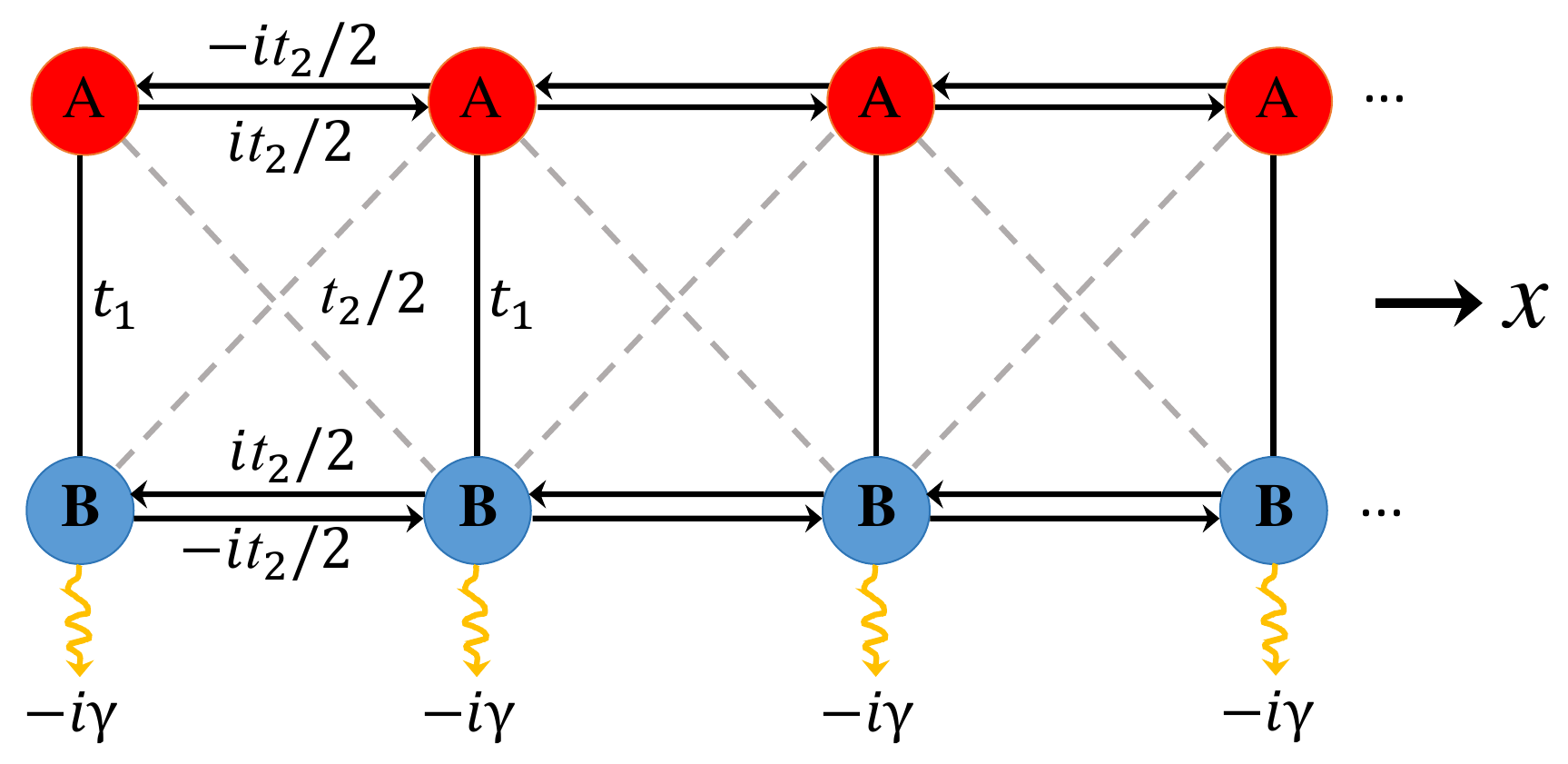}
     \caption{Schematic diagram of the non-Hermitian lattice. Each unit cell, labeled by spatial
     coordinate $x$, contains two sites A and B.
     }
     \label{fig1}
     \end{figure} 

Let's consider a one-dimensional non-Hermitian lattice (see Fig.~\ref{fig1}), from which the walker can escape during the quantum walk. The state of the
system $\left\vert \psi\right\rangle $ evolves according to the following equations of
motion (we set $\hbar=1$):%
\begin{align}
   & i\frac{d\psi_{x}^{A}}{dt} =t_{1}\psi_{x}^{B}+i\frac{t_{2}}{2}\left(  \psi
    _{x-1}^{A}-\psi_{x+1}^{A}\right)  +\frac{t_{2}}{2}\left(  \psi_{x-1}^{B}%
    +\psi_{x+1}^{B}\right)  ,\notag  \\
    &i\frac{d\psi_{x}^{B}}{dt} =t_{1}\psi_{x}^{A}-i\frac{t_{2}}{2}\left(
          \psi_{x-1}^{B}-\psi_{x+1}^{B}\right)  +\frac{t_{2}}{2}\left(  \psi_{x-1}%
          ^{A}+\psi_{x+1}^{A}\right) \notag \\
     & \ \ \ \ \ \ \ \ \ \ -i\gamma\psi_{x}^{B},
     \label{eq1}
\end{align}
where $\psi_{x}^{A}=\left\langle x,A|\psi\right\rangle $ and $\psi_{x}%
^{B}=\left\langle x,B|\psi\right\rangle $ are the amplitudes of the walker on the sublattices
A and B at the site $x$. Without loss of generality, we choose the hopping amplitude
parameters $t_{1}$ and $t_{2}$ to be real numbers. The onsite imaginary
potential $-i\gamma$ describes the loss particles on B sites with rate
$2\gamma$. This model differs from the previous quantum work model \cite{PhysRevLett.102.065703}, as it features the
NHSE. This can be seen clearly by mapping a similar model in Ref.~\cite{PhysRevLett.116.133903}, to the non-Hermitian Su-Schrieffer-Heeger (SSH)
model with nonreciprocal hopping \cite{PhysRevLett.121.086803}. The mathematical relation 
of these lattice models can be seen in Appendix.~\Ref{Appendix A}.

\begin{figure*}[tbp]
     \centering
     \includegraphics[scale=0.6]{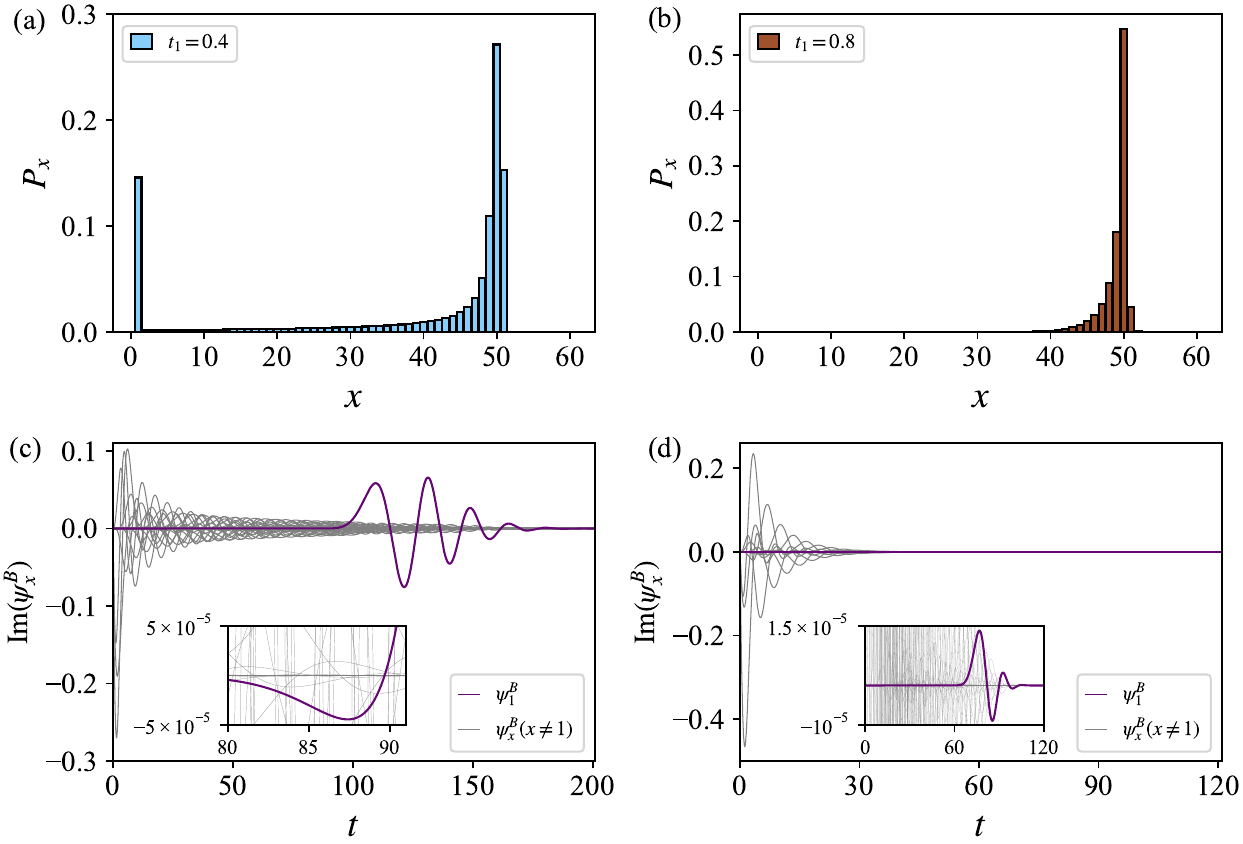}
     \caption{(a),(b) The spatially resolved loss probability $P_{x}$ for a walker initiated at $ x_{0}=50$.
     $t_{1} = 0.4$ for (a) and $t_{1}=0.8$ for (b). The chain length $L=60$. (c), (d) The corresponding 
     time evolution of $\psi_{x}^{B} $ to (a), (b) respectively. The common parameters 
     $t_{2} = 0.5$ and $\gamma = 0.8$ are fixed throughout (a)–(d).
     }
     \label{fig2}
     \end{figure*}

For a general Hamiltonian $\mathcal{\hat{H}}=\hat{H}-i\hat{\Gamma},$ where
$\hat{H}$ and $\hat{\Gamma}$ are Hermitian operators, the norm of a quantum
state $\left\vert \psi\right\rangle $ evolves according to $\frac{d}%
{dt}\left\langle \psi|\psi\right\rangle =-2\left\langle \psi\left\vert
\hat{\Gamma}\right\vert \psi\right\rangle .$ In the non-Hermitian quantum
walk model we consider, the system decays according to $\frac{d}{dt}\left\langle \psi
|\psi\right\rangle =-\sum_{x}2\gamma\left\vert \psi_{x}^{B}\right\vert ^{2},$
and the local decay probability on site $x$ is 
\begin{equation}
P_{x}=\int_{0}^{\infty}2\gamma\left\vert \psi_{x}^{B}\left(  t\right)
\right\vert ^{2}dt.
\label{eq2}
\end{equation}
If the initial state $\left\vert \psi\left(  0\right)  \right\rangle $ is
normalized, the decay probability distribution satisfies $\sum_{x}P_{x}=1$.
Now, suppose a walker starts from some sublattice A of site $x$ at time $t=0,$
namely $\psi_{x}^{A}\left(  0\right)  =\delta_{x,x_{0}},\psi_{x}^{B}\left(
0\right)  =0,$ and involves freely under the equations of motion (\ref{eq1}). 
The hop between different sites drive the walker away from $x_{0},$
and during this quantum walk, the walker can escape from any B sites. This can
be seen clearly in Fig.~\ref{fig2}(a) and \ref{fig2}(b), which is the numerical solution of $P_{x}.$ The
distribution of $P_{x}$ is left-right asymmetric and it originates from the
NSHE, the walker tends to jump towards the left as all
eigenstates are localized at the left edge.

A fascinating property of the system is the edge burst \cite{PhysRevLett.128.120401}, namely the
appearance of a prominent peak in the loss probability at the edge, with the
nearby almost invisible decaying tail (see Fig.~\ref{fig2}(a)). 
 Such an unexpected peak was numerically seen in the earlier Ref.~\cite{Wang_2021} and it was attributed to topological edge states. 
 This interpretation was regarded as wrong in \cite{PhysRevLett.128.120401} for the disappearance of the high peak in the topological nontrivial region. 
To explore how the 
walker propagates in the lattice and forms a burst loss probability peak at the edge, we focus on the time evolution of
$\psi_{x}^{B}.$ The numerical result shows that $\psi_{x}^{B} $ are purely imaginary. Furthermore, when there is the edge burst phenomenon, 
the dynamical evolution of wave function at the edge is distinct from the bulk one.
We can see this clearly in Fig.~\ref{fig2}(c) that $\psi_{1}^{B} $ has a tremendous large increased
amplitude peak after the first tiny peak, while other $\psi_{x\neq1}^{B}$ 
oscillate with a decreasing amplitude as $t$ becomes larger. 
This interesting feature is crucial to form a burst edge
peak. There is no such tremendous large increased peak for $\psi_{1}^{B} $ in Fig.~\ref{fig2}(d), where all $\psi_{x}^{B} $ have the same decreased oscillation behavior,
and the corresponding edge loss probability $P_{1}$ in Fig.~\ref{fig2}(b) is very small.

\section{\label{sec:level3} Non-Hermitian time-dependent perturbation theory}
The analytical expression of $\psi_{x}^{B}\left(  t\right) $ can be obtained via time-dependent perturbation theory.
The sketch of this theory is encapsulated as follows.
We consider a non-Hermitian Hamiltonian $\hat{H}\left(  t\right)$ such that it can be split 
into time-independent part $\hat{H}_{0}$ and time-dependent part $\hat{H}^{\prime}\left(  t\right)$, namely 
\begin{equation}
\hat{H}\left(  t\right)  =\hat{H}_{0}+\hat{H}^{\prime}\left(  t\right).
\label{eq3}
\end{equation}
The corresponding Schr\"{o}dinger equation is
\begin{equation}
i\frac{\partial}{\partial t}\left\vert \Psi\left(  t\right)  \right\rangle
=\hat{H}\left(  t\right)  \left\vert \Psi\left(  t\right)  \right\rangle .
\label{eq4}
\end{equation} 
When $\hat{H}^{\prime}\left(  t\right)  =0$, Eq.~(\ref{eq4}) can be solved if we know
the solution of eigen-equations $\hat{H}_{0}\left\vert n\right\rangle =E_{n}\left\vert n\right\rangle $ with
eigenstates $\left\vert n\right\rangle $ and eigenvalues $E_{n}$, that is 
\begin{equation}
   \left\vert \Psi\left(  t\right)  \right\rangle = e^{-i\hat{H}_{0}t}%
     \left\vert \Psi\left(  0\right)  \right\rangle ,
     \label{eq5}
\end{equation}
where $  \left\vert \Psi\left(  0\right)  \right\rangle$ is the linear combination 
of $\left\vert n\right\rangle $. If $\hat{H}^{\prime}\left(  t\right)  \neq0,$ it is no longer a stationary problem
and we are interested in the case that the initial state $\left\vert k\right\rangle $
is one of the eigenstates of $\hat{H}_{0}.$ Due to the perturbation of $\hat{H}^{\prime
}\left(  t\right),$ the state $\left\vert k\right\rangle $ is an unstable
state. We assume the system is a superposition of the eigenstates of $\hat{H}_{0}$
for $t>0$ and given by
\begin{equation}
\left\vert \Psi\left(  t\right)  \right\rangle =\sum_{n}\left\vert \psi
_{n}\left(  t\right)  \right\rangle =\sum_{n}c_{n}\left(  t\right)
e^{-iE_{n}t}\left\vert n\right\rangle .
\label{eq6}
\end{equation}
Substituting this ansatz wave function into Schr\"{o}dinger equation~%
(\ref{eq4}) and multiplying by the state $\left\langle m\right\vert$, we get a
coupled differential equation of wave function expansion coefficient under
unperturbed representation $\hat{H}_{0}$:
\begin{equation}
i\frac{dc_{m}\left(  t\right)  }{dt}=\sum_{n}H_{mn}^{\prime}\left(  t\right)
c_{n}\left(  t\right)  e^{i\left(  E_{m}-E_{n}\right)  t},
\label{eq7}
\end{equation}
where $H_{mn}^{\prime}\left(  t\right)  =$\ $\left\langle m\right\vert
\hat{H}^{\prime}\left(  t\right)  \left\vert n\right\rangle$. To solve the
differential equation (\ref{eq6})
 with the initial condition $c_{m}\left(  0\right)  =\delta_{mk}$, we take the
perturbation expansion of $c_{m}\left(  t\right)  $ and use the iteration method.
Specifically, $c_{m}\left(  t\right)  $ can be written as%
\begin{equation}
c_{m}\left(  t\right)  =c_{m}^{\left(  0\right)}
+c_{m}^{\left(  1\right)  } +c_{m}^{\left(  2\right)}  +\cdots,
\label{eq8}
\end{equation}
where $c_{m}^{\left(  1\right)  } , ~ c_{m}^{\left(  2\right)}  ,~\cdots$ signify 
amplitudes of the first order, second order, and so on in the strength parameter of 
time-dependent Hamiltonian. Plugging Eq.~(\ref{eq8})
 into Eq.~(\ref{eq7}) and comparing the order of perturbation, we can get the $l$ th-order equation:
\begin{equation}
i\frac{d}{dt}c_{m}^{\left(  l\right)  }\left(  t\right)  =\sum_{n}%
H_{mn}^{\prime}\left(  t\right)  c_{n}^{\left(  l-1\right)  }\left(  t\right)
e^{i\left(  E_{m}-E_{n}\right)  t}.
\label{eq9}
\end{equation}
This equation can be solved easily by integrating it directly if we know all of $(l-1)$ th-order solutions 
$ c_{n}^{\left(  l-1\right)}$. Thus, we can solve Eq.~(\ref{eq7}) step by step, and the solution up to $l$ th-order is
\begin{equation}
c_{m}\left(  t\right)  \simeq c_{m}^{\left(  0\right)  }\left(  t\right)
+c_{m}^{\left(  1\right)  }\left(  t\right)  +\cdots+c_{m}^{\left(  l\right)
}\left(  t\right)  .
\label{eq10}
\end{equation}
It should be noticed that this approximate solution has a convergent radius
$t_{0}$, which is related to the perturbation expansion order and the strength parameters of perturbation.

\section{\label{sec:level4} Burst peak from time-dependent perturbation theory}
\begin{figure*}[tbp]
    \centering
   \includegraphics[scale=0.55]{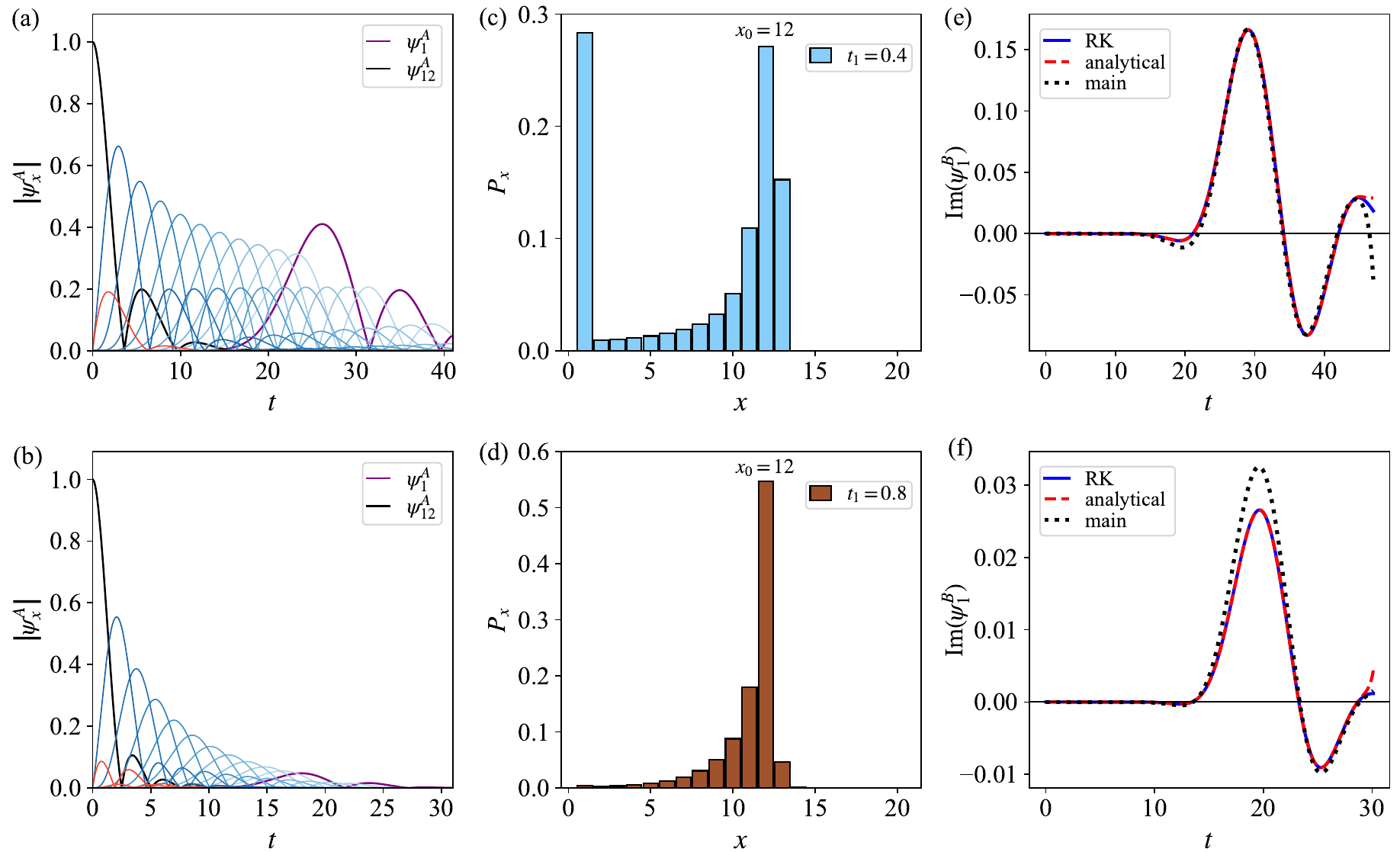}
    \caption{(a), (b) The modulus of the wave functions $\psi_{x}^{A}\left(
       t\right)$ for a walker initiated at $ x_{0}=12$.
    $t_{1} = 0.4$ for (a) and $t_{1}=0.8$ for (b). The gradient blue (red) curves 
    represent the sites that are on the left (right) side of the initial site, namely from $| \psi_{11}^{A}(t)\vert $  to  $| \psi_{2}^{A}(t)\vert $ ($| \psi_{13}^{A}(t)\vert $  to  $| \psi_{20}^{A}(t)\vert $). 
    The lighter the blue (red) color of the curve is, the further the site is away from the initial site.  Specifically, the black curve represents the initial 12A site while the purple 
    curve represents the left edge site 1A. The chain length $L=20$. (c), (d) The corresponding 
     spatially resolved loss probability $P_{x}$ to (a), (b) respectively. (e) and (f) are the edge wave functions  $\psi_{1}^{B}\left(
       t\right)$. The blue solid line means the Runge-Kutta numerical solution. The evaluation of both analytical (red dashed) and main paths (black dotted) are up to 90th-order perturbation. The common parameters 
    $t_{2} = 0.5$ and $\gamma = 0.8$ are fixed throughout (a)–(f).
    }
    \label{fig3}
\end{figure*} 
In the non-Hermitian quantum walk model, the onsite potential operator is set
as the unperturbed Hamiltonian $\hat{H}_{0}$. Its real-space matrix
elements are $\left\langle x,s\right\vert \hat{H}_{0}\left\vert x^{\prime
},s^{\prime}\right\rangle =-\frac{i\gamma}{2}\delta_{xx^{\prime}}%
\delta_{ss^{\prime}}\left[  1-\left(  \sigma_{z}\right)  _{ss^{\prime}%
}\right]$, where $x,x^{\prime}$ referring to the location of the unit cell and
$s,s^{\prime}=A,B$ referring to the sublattice label. The eigen-equation is $\hat
{H}_{0}\left\vert x,s\right\rangle =E_{x}^{s}\left\vert x,s\right\rangle $,
with two $N$-fold eigenvalues $E_{x}^{A}=0$ and $E_{x}^{B}=-i\gamma$ respectively. The
hopping of the walker is treated as the perturbation operator $\hat{H}^{\prime}$ and
its matrix elements are $H_{xs,x^{\prime}s^{\prime}}^{\prime}=\left\langle
x,s\left\vert \hat{H}^{\prime}\right\vert x^{\prime},s^{\prime}\right\rangle
$. Thus, according to the perturbation procedure in Sec.~\ref{sec:level3}, we 
can evaluate the evolution of the amplitude of the walker on any site.
Specifically, Eq.~(\ref{eq7}) will be reduced to
\begin{equation}
i\frac{dc_{x}^{s}\left(  t\right)  }{dt}=\sum_{x^{\prime}s^{\prime}%
}H_{xs,x^{\prime}s^{\prime}}^{\prime} c_{x^{\prime}%
}^{s^{\prime}}\left(  t\right)  e^{i\left(  E_{x}^{s}-E_{x^{\prime}%
}^{s^{\prime}}\right)  t}.
\label{eq11}
\end{equation}
One can integrate both sides of this equation
 and get the formal solution,%
\begin{equation}
c_{x}^{s}\left(  t\right)  =-i\sum_{x^{\prime}s^{\prime}}\int_{0}%
^{t}H_{xs,x^{\prime}s^{\prime}}c_{x^{\prime}}^{s^{\prime}}\left(  t^{\prime
}\right)  e^{i\left(  E_{x}^{s}-E_{x^{\prime}}^{s^{\prime}}\right)  t^{\prime
}}dt^{\prime}.
\label{eq12}
\end{equation}
with the relation $\psi_{x}^{s}\left(  t\right)  =e^{-iE_{x}^{s}t}c_{x}%
^{s}\left(  t\right)  ,$ the formal solution can be written as the relation
between the amplitude of the different sites:
\begin{equation}
\psi_{x}^{s}\left(  t\right)  =-ie^{-iE_{x}^{s}t}\sum_{x^{\prime}s^{\prime}%
}\int_{0}^{t}e^{iE_{x}^{s}t^{\prime}}H_{xs,x^{\prime}s^{\prime}}%
\psi_{x^{\prime}}^{s^{\prime}}\left(  t^{\prime}\right)  dt^{\prime}.
\label{eq13}
\end{equation}
 Initial conditions $\psi_{x}^{A}\left(  0\right)  =\delta_{x,x_{0}}$
and $\psi_{x}^{B}\left(  0\right)  =0 $ guarantee that the amplitudes $\psi
_{x}^{A}\left(  t\right)  $ remain real and $\psi_{x}^{B}\left(
t\right)  $ remain purely imaginary for all $t$. This follows from the perturbation analysis (see details in Appendix.~\Ref{Appendix B}). Alternatively, one can check this conclusion through the iteration
equation (\ref{eq13}). For example, if the sublattice label $s=A,$ the nonzero factors
$H_{xs,x^{\prime}s^{\prime}}\psi_{x^{\prime}}^{s^{\prime}}\left(  t^{\prime
}\right)  $ are imaginary under the assumption that $\psi_{x^{\prime}%
}^{s^{\prime}}\left(  t^{\prime}\right)  $ are real for $s^{\prime}=A$ and
imaginary for $s^{\prime}=B.$ The amplitudes $\psi_{x}^{A}\left(  t\right)  $
thus keep real, which is a self-consistent result for equation (\ref{eq13}).

For concreteness, we consider a $20$-site lattice with the walker initially prepared on sublattice $A$ of site $12$. 
 Analytical expressions of wave functions $\psi_{x}^{s}\left(  t\right)$ are obtained using
time-dependent perturbation theory, and these expressions agree well with the numerical results at
the desired time. It finds that the dynamical behavior of the walker
along the $A$ chain is crucial to forming a remarkable loss peak at the edge.
Increasing the hopping parameter $t_{1}$ strengthens the coupling between chains $A$ 
and $B$ for fixed $t_{2}$ and $\gamma$,  thereby accelerating walker 
dissipation and consequently shortening its travel time on chain A. This manifests in 
Fig.~\ref{fig3}(a) and \ref{fig3}(b), where $\psi_{x}^{A}\left(  t\right)$ 
exhibits smaller amplitudes and faster convergence to zero for larger $t_{1}$ upon the walker's arrival at $x$. 
Another perspective to see this is that the dissipation probabilities $P_{x}$ near 
the initial location is larger as $t_{1}$ increases (see Fig.~\ref{fig3}(c) and \ref{fig3}(d)).
If the system satisfies the imaginary gapless condition $t_{1} < t_{2}$, 
the walker will decay slowly with the algebraic behavior of bulk $P_{x}$ \cite{PhysRevLett.128.120401,PhysRevB.103.L241408}, 
which leads a large amplitude of $\psi_{1}^{A}$ (see Fig.~\ref{fig3}(a)).
On the other hand, this non-Hermitian lattice model features the NHSE, namely
the exponential localization of all eigenstates at the edge, which is
characterized by the generalized Bloch factor $\beta$ with $\left\vert
\beta\right\vert =\sqrt{\left\vert \left(  t_{1}-\gamma/2\right)  /\left(
t_{1}+\gamma/2\right)  \right\vert }$. In the case $\left\vert \beta
\right\vert <1$ for $t_{1}>0,$ all skin modes are localized at the left edge. The
NHSE induces leftward walking along the $A$ chain with $max[| \psi_{x<x_0}^{A} \vert]$ much larger 
than $max[| \psi_{x>x_0}^{A} \vert]$, as shown in Fig.~\ref{fig3}(a) and \ref{fig3}(b).
The walker becomes trapped at the left edge once it arrives at $A_{1}$, which leads to the Rabi-like
oscillation between sublattice $A_{1}$ and $B_{1}$. With such a Rabi-like oscillation picture in mind, 
one can easily combine the prominent burst of the purple $| \psi_{1}^{A}(t) \vert $ curve in Fig.~\ref{fig3}(a) 
with the dissipation probability burst in $B_{1}$ in Fig.~\ref{fig3}(c). 
A larger $| \psi_{1}^{A}(t) \vert $ will result in a larger amplitude of 
the quantum jump $|1,A \rangle \rightarrow   |1,B \rangle$.  

The complete analytical expression of $\psi_{x}^{s}\left(  t\right)  $ can be
interpreted as the sum of all physically allowed paths the walker traverses from
 $\left\vert x_{0},A\right\rangle $ to  $\left\vert
x,s\right\rangle $ during time $t$. For example, the walker can reach $\left\vert x_{0},B\right\rangle $ via only one single-step quantum jump from $\left\vert x_{0},A\right\rangle $, 
while four distinct paths exist for a two-step jump. These paths correspond to the first- and 
second-order perturbation contributions in $\psi_{x_{0}}^{B}\left(  t\right)$,
respectively. We can classify every perturbation process by its final-step quantum jump. 
The sum of all perturbation terms with the same final-step quantum jump is 
the total transition amplitude from a certain nearest-neighbor site of 
$\left\vert x,s\right\rangle $ to $\left\vert x,s\right\rangle $.
This is the physical interpretation of the integral formula (\ref{eq13}).
When the walker travels from $\left\vert x_{0},A\right\rangle $ to $\left\vert
    x,s\right\rangle $ for a bulk site $x$, there are five transition
    process, corresponding to five final step quantum jumps
    $\left\vert x^{\prime},s^{\prime}\right\rangle \rightarrow\left\vert
    x,s\right\rangle $ with $s\prime=A,B$ for $x^{\prime}=x\pm1$ and $s^{\prime
    }\neq s$ for $x^{\prime}=x$. However, the number of different final step
    quantum jumps is reduced to three if the walker arrives at $x=1$. This feature
    leads the nonzero hopping matrix element $H_{xs,x^{\prime}s^{\prime}}$ in
    equation (13) to be five for the bulk $x$ and three for the edge one.
Thus, the contrasting evolution of
 bulk and edge wave functions in the edge burst phenomenon
originates from
 their distinct nearest-neighbor site configurations. These configurations constrain the permissible  
paths a walker can travel from the initial state. For example, a quantum jump process
$\left\vert x+1,A\right\rangle \rightarrow\left\vert x,B\right\rangle
\rightarrow\left\vert x-1,A\right\rangle \rightarrow\left\vert x,B\right\rangle$ 
is allowed in the bulk sites but forbidden in the edge site.
 Another perspective to see this is that if we neglect the back transition from the two forward nearest-neighbor sites, the
bulk wave function evolves like the edge wave function as it can be viewed as the new
physical edge artificially. Furthermore, we find that the edge wave function
$\psi_{1}^{B}\left(  t\right)  $ can be approximated by the interference of
transition amplitude from two adjacent non-decay sites $A_{1}$ and $A_{2}$, or
symbolically,%
\begin{equation}
    \psi_{1}^{B}\left(  t\right)  \simeq-ie^{-\gamma t}\int_{0}^{t}e^{\gamma
    t^{\prime}}\left[  t_{1}\psi_{1}^{A}\left(  t^{\prime}\right)  +\frac{t_{2}%
    }{2}\psi_{2}^{A}\left(  t^{\prime}\right)  \right]  dt^{\prime}.
\end{equation}
Fig.~\ref{fig3}(e) and ~\ref{fig3}(f) visually confirm this, as we viewed 
the transition paths with final step quantum jumps $A_{1}$ to $B_{1}$ or $A_{2}$ to $B_{1}$ as 
the main paths, which fit well with analytical and Runge-Kutta numerical results.  The reason is that the walker mainly propagates along the $A$ chain and the transition 
amplitude from the $B_{2}$ site can be neglected as $\psi_{2}^{B}\left(  t\right)  $ is very small.
Compared with the previously mentioned Rabi-like oscillation between
 $A_{1}$ and $B_{1}$, a more accurate picture can be drawn now. The walker starts from 
 the initial site and propagates with a preference to the left due to the NHSE. 
 After its arrival on site 2, the walker oscillates in an approximately closed loop 
 formed by $A_{1}$, $A_{2}$ and $B_{1}$ until the particle finally decays out 
 in $B_{1}$. $B_{2}$ can be excluded out of the loop since  $\psi_{2}^{B}\left(  t\right)  $ is 
 very small. In the edge burst region, both $\psi_{1}^{A}\left(  t\right)  $ and 
 $\psi_{2}^{A}\left(  t\right)  $ are relatively large as demonstrated in Fig.~\ref{fig3}(a). 
 Whereas they are fairly small in the region without edge burst as shown in Fig.~\ref{fig3}(b).

 The above discussions are all based on the initial condition that the walker 
 starts from some sublattice A. However, we note that the initial condition is also crucial for the 
 formation of the prominent loss peak at the edge. One can easily check that the edge loss peak is 
 too small to be observed for the initial case that the walker starts from sublattice B even though 
 the system features the NSHE and imaginary gap closing. See the analysis of the effect of the initial site in Appendix.~\Ref{Appendix C}.

\begin{figure*}
    \includegraphics[scale=0.53]{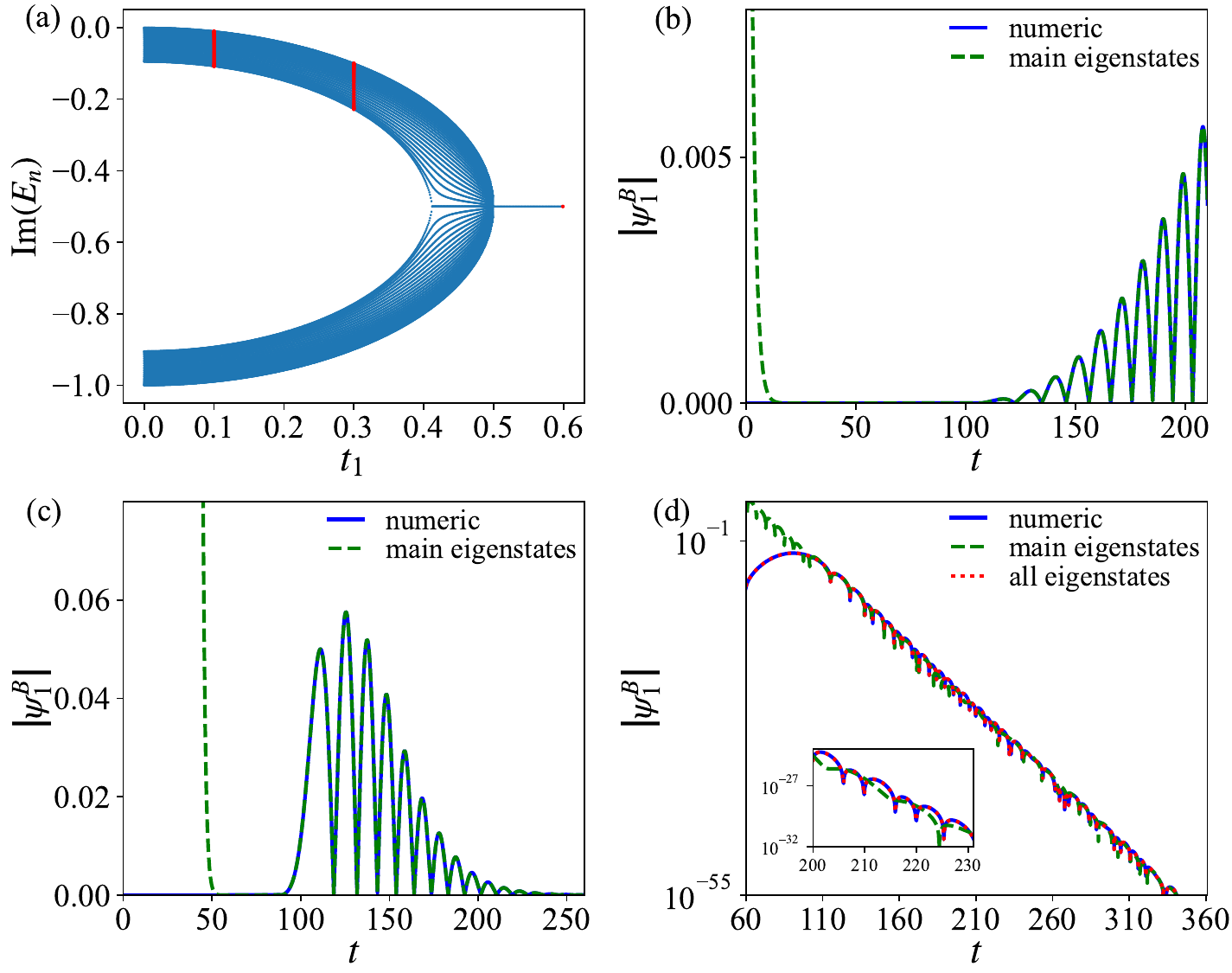}
    \caption{(a) The imaginary part of energy $E_{n}$ with $L=60$. 
    The red points at $t_{1}=0.1,t_{1}=0.3$, and $t_{1}=0.6$ correspond to the eigenstates 
    we choose to simulate the time evolution of $|\psi^{B}_{1}(t)|$. 
    Other parameters are $t_{2}=0.4,\gamma=1$ and the walker is initially put 
    at site 40A. (b) The green dashed line represents 
    the simulation by choosing 60 eigenstates with relatively 
    large imaginary parts of eigenvalues. The blue solid line represents the Runge-Kutta numeric 
    solution. $t_{1}=0.1$, other parameters are the same as (a). (c) Similar to (b) except 
    that $t_{1}=0.3$. (d) Similar to (b) except that $t_{1}=0.6$. 
    The red dashed line represents the simulation of all 
    120 eigenstates, which exactly reveal the Runge-Kutta numeric solution.     }
    \label{fig4}
\end{figure*}
\section{\label{sec:level5}The Eigen-modes and the edge wave function}
To investigate the role that different eigenstates play in the evolution process, we decompose 
the non-Hermitian Hamiltonian by the biorthogonal bases
\begin{align}
\hat{H}=\sum_{n} E_{n}|n_{R}\rangle \langle n_{L}|,
\end{align}
where $|n_{R}\rangle$ and $\langle n_{L}|$ are the right and left eigenstates, 
respectively, satisfying biorthogonality $\langle m_{L}|n_{R} \rangle=\delta_{mn}$
and completeness $\sum_{n}|n_{R}\rangle \langle n_{L}|=\mathbb{I} $.  
For the non-Hermitian lattice we investigate, the state of the system evolves according to
\begin{align}
    |\psi(t)\rangle=e^{-i\hat{H}t} |\psi(0)\rangle, \label{evolution}
\end{align}
where the initial state $|\psi(0)\rangle$ can be expressed by a superposition of $|n_{R}\rangle$
\begin{align}
    |\psi(0)\rangle=\sum_{n}a_{n}|n_{R}\rangle,
\end{align}
where $a_{n}$ is determined by the linear equations
\begin{align}
    \psi^{s}_{x}(0)=\sum_{n}a_{n}\phi^{s}_{n,x},
\end{align}
with $\phi^{s}_{n,x}\equiv \langle x,s|n_{R}\rangle$. Thus, we can rewrite the evolution equation (\ref{evolution}) as 
\begin{align}
    |\psi(t)\rangle=\sum_{n} e^{-iE_{n}t}a_{n}|n_{R}\rangle. \label{evolution2}
\end{align}
Since $E_{n}={\rm Re}(E_{n})+i{\rm Im}(E_{n})$, we can rewrite the evolution equation (\ref{evolution2}) as
\begin{align}
    |\psi(t)\rangle=\sum_{n} e^{-i{\rm Re}(E_{n})t} e^{{\rm Im}(E_{n}) t} a_{n}|n_{R}\rangle. \label{evolution3}
\end{align}
Focusing on evolution equation (\ref{evolution3}),  the long time behavior of $|\psi(t)\rangle$ 
is determined by the eigenstates with relatively large imaginary parts of 
 eigenvalues since the exponential factor in front of them decay more slowly. 
 So the imaginary parts of the eigenvalues play an important role in the time evolution of the wave function. 
 Specifically, one can verify that the imaginary part of any eigenvalue of the system is always 
 negative owing to the dissipation of the whole system. Moreover, the imaginary part of the eigenvalues 
 are twofold degenerate and symmetric to $-\frac{i}{2}\gamma$, which is demonstrated in Fig.~\ref{fig4} (a). 
 The reason is that the Hamiltonian of the quantum walk $\hat{H}$ can be divided 
 into two parts $\hat{H}=\hat{H}_{1}-\frac{i\gamma}{2}\hat{I}$ with the first part $\hat{H}_{1}$ satisfying 
 chiral symmetry and parity-time symmetry \cite{PhysRevLett.116.133903}. More details about the symmetry analysis of 
 the Hamiltonian are given in Appendix.~\Ref{Appendix A}.
 
We still take a 60-site lattice as an example to show 
the role of the imaginary part of the eigenvalues play in the time evolution of 
the wave function. By picking 60 eigenstates with relatively 
large imaginary part of eigenvalues out of all 120
eigenstates, we fit the evolution of the wave function $|\psi^{B}_{1}(t)|$ and 
show its result in Fig.~\ref{fig4}(b-d). For $t_{1}=0.1$, as one can see from 
the green dashed line in Fig.~\ref{fig4}(b), these 60 
eigenstates can describe the evolution of $|\psi^{B}_{1}(t)|$ well as early as 
$t=15$. This is because the other 60 
eigenstates have much smaller imaginary parts of eigenvalues and their contribution to 
the wave function vanishes quickly. While in Fig.~\ref{fig4}(c) with a different 
$t_{1}=0.3$, these 60 eigenstates can fit the analytical 
solution well only after $t=50$, which is due to the 
non-negligible contribution of other 60 eigenstates before 
$t=50$. When $t_{1}=0.6$, all eigenstates have the same 
imaginary part of the eigenvalues, which indicates that they contribute 
equally to the decay behavior of the wave function. If we still choose half of all eigenstates 
to fit the evolution of the wave function, the accurate simulation would fail even at 
a long-evolution time. Although the order of magnitude of the wave function can be well 
described by half of the eigenstates after $t=110$, the oscillation details are missed 
due to the exclusion of the other half of the eigenstates, which is demonstrated clearly 
in the green dashed line in the subfigure of Fig.~\ref{fig4}(d). Apparently, the combination of 
all eigenstates can accurately restore the evolution of the wave function, which is shown in the 
red dashed line in Fig.~\ref{fig4}(d).

\begin{figure}
    \includegraphics[scale=0.4]{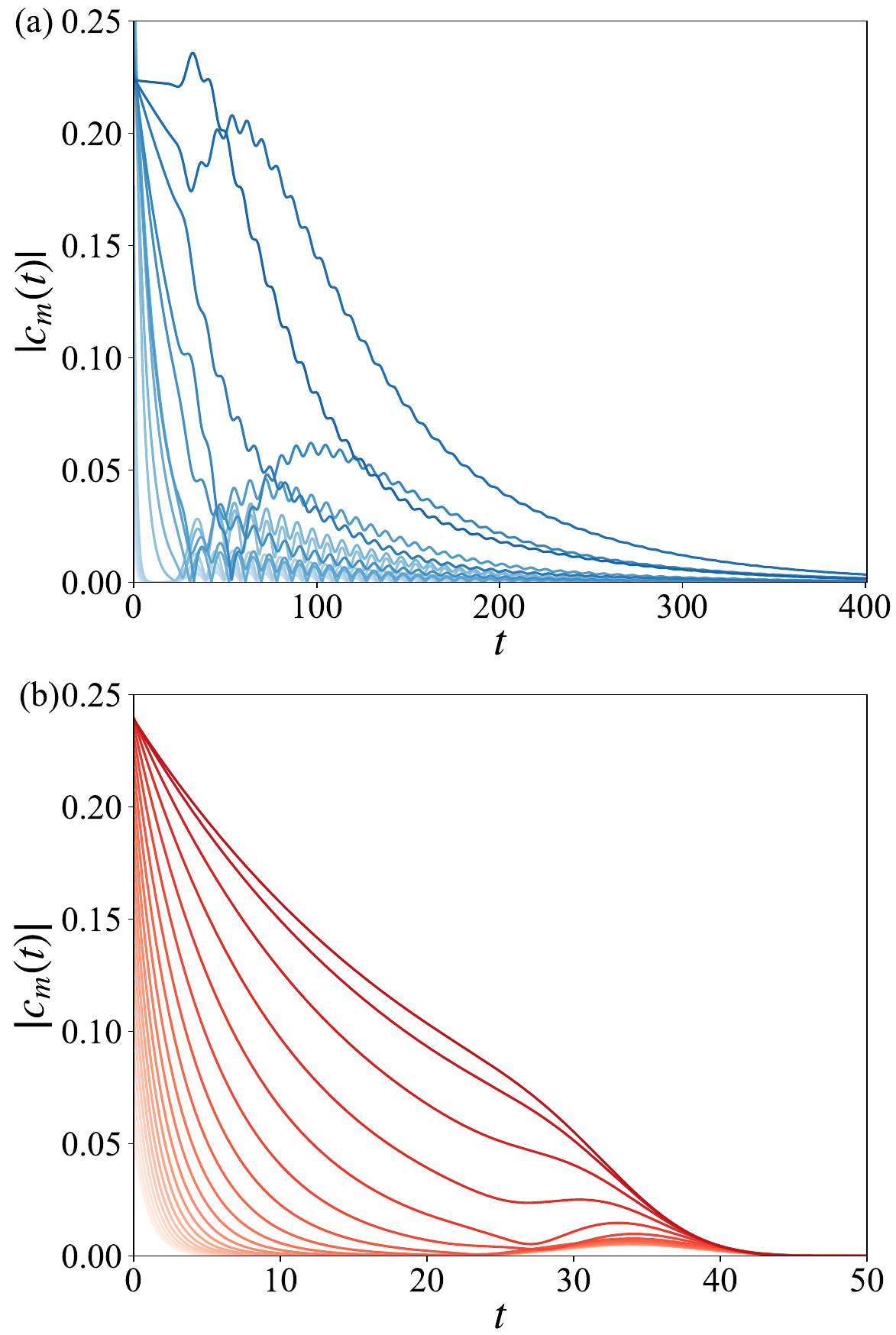}
    \caption{The PBC eigenstates weight $| c_{m}(t) \vert $ versus $t$. 
    The gradient color represents different $ \{ | c_{m}(t) \vert \}$ with different eigenvalues. 
    The darker the color of the curve is, the larger the imaginary part of the eigenvalue is. 
    One can tell that $| c_{m}(t) \vert $ with a larger imaginary part of the eigenvalue has 
    a larger integral over $t$ for both (a) and (b). (a): The edge burst region. 
    $t_{1}=0.1, t_{2}=0.4, \gamma=1$, the initial walker was put in 15A for a 20-site lattice. 
    (b): The region without edge burst. Similar to (a) except for $t_{1}=0.6$.  }
    \label{fig5}
\end{figure}

The above discussions are all based on the imaginary part of the open boundary 
condition (OBC) spectrum $\{E_{n}\}$. Another perspective is to investigate the imaginary part of the 
periodic boundary condition (PBC) spectrum $\{E^{'}_{m}\}$, which differs greatly from 
the OBC one due to the NHSE. For the dynamical evolution of the OBC system, 
one can also expand the wave function in the eigenstates of its PBC counterpart through a transform matrix $T$,
\begin{align}
    |\psi(t)\rangle =\sum_{m} c_{m}(t) |m_{k}\rangle
\end{align}
where $\{|m_{k}\rangle\}$ are the right eigenstates of the PBC Hamiltonian $H_{PBC}$ with eigenvalue $E^{'}_{m}$. One can 
easily find that the time dependent coefficient $c_{m}(t)$ satisfies $c_{m}(t)=\sum_{n} T_{nm} e^{-i E_{n}t}a_{n}$. 
Obviously, the module of $c_{m}(t)$ reveals the weight of the eigenstate $|m_{k}\rangle$ plays in the dynamical evolution. 
To investigate the relationship between the weight $| c_{m}(t) \vert $ and the imaginary part of the eigenvalue 
${\rm Im} (E^{'}_{m})$, we plot the change of $| c_{m}(t) \vert $ over time in Fig.~\ref{fig5} for a 20-site system 
with the walker initialized at 15A. We have arranged $| c_{m}(t) \vert $ according to the magnitude of  ${\rm Im} (E^{'}_{m})$ 
in Fig.~\ref{fig5}. The darker the color of the curve $| c_{m}(t) \vert $ is, the larger  ${\rm Im} (E^{'}_{m})$  is. 
One can easily tell that a larger ${\rm Im} (E^{'}_{m})$ leads to $| c_{m}(t) \vert $ decays more slowly, which results 
in a larger integral area over $t$. The integral area of $| c_{m}(t) \vert $ over $t$ can be viewed as another indicator 
of the weight of different eigenstates plays in the evolution process. By analyzing an OBC dynamical evolution problem with 
the PBC eigenstates and eigenvalues, we emphasize the importance of the imaginary part of the eigenvalues 
again, no matter whether the eigenvalues are OBC's or PBC's. Edge burst is an astonishing phenomenon 
in OBC. Herein, using the PBC quantity $| c_{m}(t) \vert $ to investigate an OBC dynamical problem may 
have some connection with the imaginary gap closing in Ref.~\cite{PhysRevLett.128.120401}, where the authors 
also set a PBC criterion of an OBC phenomenon.  
\section{\label{sec:level6} Summary and discussion}
In this paper, we study the real-space dynamical evolution of quantum particles in the lossy lattices. We find the edge burst phenomenon
is closely related to the distinct evolution features between the edge wave function and bulk wave functions by a numerical simulation. We then give a sketch of 
time-dependent perturbation theory for a non-Hermitian Hamiltonian and
evaluate the analytical expression of the quantum walk wave function. 
Through the analysis of the perturbation solution,
we find the walker mainly propagates nonreciprocally along the non-decay chain as the non-Hermitian lattices feature the NHSE. Moreover,
the different evolution behaviors between the edge
and bulk can be attributed to their nearest-neighbor site configurations, which limit the 
possible path that the walker can travel from the initial site. Besides, it finds
that the main contribution to the edge wave function results from the
interference transition of the two nearest-neighbor sites. Furthermore, the
numerical diagonalization shows that the walker is mainly propagated by a
group of eigen-modes that have a relatively large imaginary part. 
 
Our work gives an explicit illustration of the edge burst
phenomenon in real space directly and provides an alternative method to study
this kind of non-Hermitian dynamical problem, for example, the quantum walk 
with a non-uniform loss rate \cite{PhysRevB.107.L140302} (details see Appendix.~\Ref{Appendix D}). 

\begin{acknowledgments}
The authors would like to thank Yumin Hu, Zheng Fan, and Yuting Lei for their helpful discussion. This work is supported by the National Natural Science Foundation of China under Grants No.~11974205, and No.~61727801, and the Key Research and Development Program of Guangdong province (2018B030325002).
\end{acknowledgments}

\begin{appendix}

\section{Mathematical relation between several lattice models}\label{Appendix A}
In Appendix A, We mainly discuss the mathematical relation 
of lattice models mentioned in Sec.~\ref{sec:level2}. The Hamiltonian of the quantum walk model with $N$ 
unit cell is%
\begin{equation}
 \hat{H}=\hat{H}_{1}-\frac{i\gamma}{2}I_{2N\times2N},
\end{equation}
where $\hat{H}_{1}$ has chiral symmetry and parity-time ($\mathcal{PT}$)
symmetry \cite{PhysRevLett.116.133903}. The chiral symmetry is defined by $\Gamma=\bigoplus\nolimits_{n}%
\sigma_{y}^{n},\Gamma\hat{H}_{1}\Gamma=-\hat{H}_{1}$ and the $\mathcal{PT}$
symmetry is defined by $\mathcal{P}=\bigoplus\nolimits_{n}\sigma_{y}%
^{n},\mathcal{T}i\mathcal{T}=-i,\mathcal{PT}\hat{H}_{1}\mathcal{T}%
^{-1}\mathcal{P}^{-1}=\hat{H}_{1}$.
The Hamiltonian $\hat{H}_{1}$ can be transformed to the non-Hermitian
SSH model with left-right asymmetric hopping by a $\frac{\pi
}{2}$ rotation about the $x$-axis to each spin if we viewed every sublattice
as a pseudo spin, or symbolically,
\begin{equation}
\hat{H}_{2}=\mathcal{R}^{-1}\hat{H}_{1}\mathcal{R},
\end{equation}
where%
\begin{equation}
\hat{H}_{2}=\left[
\begin{array}
[c]{ccccc}%
0 & t_{1}+\frac{\gamma}{2} & 0 & 0 & \cdots\\
t_{1}-\frac{\gamma}{2} & 0 & t_{2} & 0 & \cdots\\
0 & t_{2} & 0 & t_{1}+\frac{\gamma}{2} & \cdots\\
0 & 0 & t_{1}-\frac{\gamma}{2} & 0 & \ddots\\
\vdots & \vdots & \vdots & \ddots & \ddots
\end{array}
\right]  _{2N\times2N},
\end{equation}
and the spin rotate operator is $
\mathcal{R=}%
{\displaystyle\bigoplus\nolimits_{n}}
e^{-i\frac{\pi}{4}\sigma_{x}^{n}}.
$
The Hamiltonian $\hat{H}_{2}$ can also be related to the stand standard SSH
model via a similarity transformation \cite{PhysRevLett.121.086803}, or symbolically,
\begin{equation}
\hat{H}_{3}=S^{-1}\hat{H}_{2}S,
\end{equation}
where
\begin{equation}
\hat{H}_{3}=\left[
\begin{array}
[c]{ccccc}%
0 & t_{1}^{\prime} & 0 & 0 & \cdots\\
t_{1}^{\prime} & 0 & t_{2}^{\prime} & 0 & \cdots\\
0 & t_{2}^{\prime} & 0 & t_{1}^{\prime} & \cdots\\
0 & 0 & t_{1}^{\prime} & 0 & \ddots\\
\vdots & \vdots & \vdots & \ddots & \ddots
\end{array}
\right]  _{2N\times2N},
\end{equation}
and $S$ is a diagonal matrix with%
\begin{equation}
S=%
{\displaystyle\bigoplus\nolimits_{n}}
S_{n},S_{n}\equiv\beta^{n-1}\left[
\begin{array}
[c]{cc}%
1 & 0\\
0 & \beta
\end{array}
\right]  .
\end{equation}
Here the parameter is $\beta=\sqrt{\left(  t_{1}-\gamma/2\right)  /\left(
t_{1}+\gamma/2\right)  },t_{1}^{\prime}=\sqrt{t_{1}^{2}-\gamma^{4}/4}%
,t_{2}^{\prime}=t_{2}.$ Since the spin rotation and similarity transformation
does not change eigenvalues, then $\hat{H}_{1},\hat{H}_{2}$ and $\hat{H}_{3}$
share the same eigen-spectrum. Thus, for an eigen-state $\left\vert \Psi
_{3}\right\rangle =\left(  \psi_{1,A},\psi_{1,B},\cdots,\psi_{N,A},\psi
_{N,B}\right)  ^{T}$ of $\hat{H}_{3}$, there are corresponding eigenstate
$\left\vert \Psi_{2}\right\rangle =S\left\vert \Psi_{3}\right\rangle $ of
$\hat{H}_{2}$ and $\left\vert \Psi_{1}\right\rangle =\mathcal{R}\left\vert
\Psi_{2}\right\rangle $ of $\hat{H}_{1}$ both exponentially localized at an
end of the chain when $\gamma\neq0$, namely features the non-Hermitian skin
effect. The Hamiltonian of another related quantum walk model in Ref \cite{PhysRevLett.102.065703} is
\begin{equation}
\hat{H}_{4}=\hat{H}_{3}+\frac{i\gamma}{2}%
{\displaystyle\bigoplus\nolimits_{n}}
\sigma_{z}^{n}-\frac{i\gamma}{2}I_{2N\times2N}.
\end{equation}
 This model does not have edge burst phenomenon due to the lack of
non-Hermitian skin effect.

\section{Prove of wave function real-imaginary property}\label{Appendix B} 
In Appendix B, we will prove that $\psi_{x}^{A}\left(  t\right)  $ are real
and $\psi_{x}^{B}\left(  t\right)  $ are purely imaginary under the initial
condition $\psi_{x}^{A}\left(  0\right)  =\delta_{x,x_{0}},\psi_{x}^{B}\left(
0\right)  =0$. Since $\psi_{x}^{s}\left(  t\right)  =e^{-iE_{x}^{s}t}c_{x}%
^{s}\left(  t\right)  $ and $e^{-iE_{x}^{s}t}$ is real$,$we only need to
consider $c_{x}^{s}\left(  t\right)  $ and the recursion equation is%
\begin{equation}
i\frac{dc_{x}^{s}\left(  t\right)  }{dt}=\sum_{x^{\prime}s^{\prime}%
}H_{xs,x^{\prime}s^{\prime}}^{\prime}c_{x^{\prime}}^{s^{\prime}}\left(
t\right)  e^{i\left(  E_{x}^{s}-E_{x^{\prime}}^{s^{\prime}}\right)  t}.
\label{B1}
\end{equation}
The $l$-th order recursion perturbation equation is%
\begin{equation}
\frac{dc_{x}^{s\left(  l\right)  }}{dt}=-i\sum_{x^{\prime}s^{\prime}%
}H_{xs,x^{\prime}s^{\prime}}^{\prime}c_{x^{\prime}}^{s^{\prime}\left(
l-1\right)  }e^{i\left(  E_{x}^{s}-E_{x^{\prime}}^{s^{\prime}}\right)  t}.
\label{B2}
\end{equation}
For $l=0,$ we have $c_{x}^{A}=\delta_{x,x_{0}},c_{x}^{B}=0$ and the first-order
perturbation $c_{x}^{s\left(  1\right)  }$ can be acquired from Eq.~(\ref{B2}). We
note that $c_{x\pm1}^{A\left(  1\right)  }$ are real and $c_{x\pm1}^{B\left(
1\right)  },c_{x}^{B\left(  1\right)  \text{ }}$ are imaginary. If we suppose
all nonzero $c_{x}^{A\left(  l-1\right)  }$ are real and $c_{x}^{B\left(
l-1\right)  }$ are imaginary, the $c_{x}^{A\left(  l\right)  }$ then can be
deduced to be real. The argument is as follows, the matrix elements
$H_{xA,x^{\prime}A}^{\prime}$ are imaginary and $H_{xA,x^{\prime}B}^{\prime}$
are real, the combination $-H_{xs,x^{\prime}s^{\prime}}^{\prime}c_{x^{\prime}%
}^{s^{\prime}\left(  l-1\right)  }e^{i\left(  E_{x}^{s}-E_{x^{\prime}%
}^{s^{\prime}}\right)  t}$ thus always to be real. From Eq.~(\ref{B2}), we
conclude that $c_{x}^{A\left(  l\right)  }$ are real. Add all perturbation order
of $c_{x}^{A\left(  l\right)  }$ together, we have
\begin{equation}
c_{x}^{A}=c_{x}^{A\left(  0\right)  }+c_{x}^{A\left(  1\right)  }+\cdots
+c_{x}^{A\left(  l\right)  }+\cdots.
\label{B3}
\end{equation}
Thus, $\psi_{x}^{A}\left(  t\right) $ is proved to be real. We can also prove
$\psi_{x}^{B}\left(  t\right)  $ are purely imaginary by a similar procedure.

\section{The effect of the initial site on edge burst}\label{Appendix C} 
\begin{figure}[tbp]
    \centering
    \includegraphics[scale=0.45]{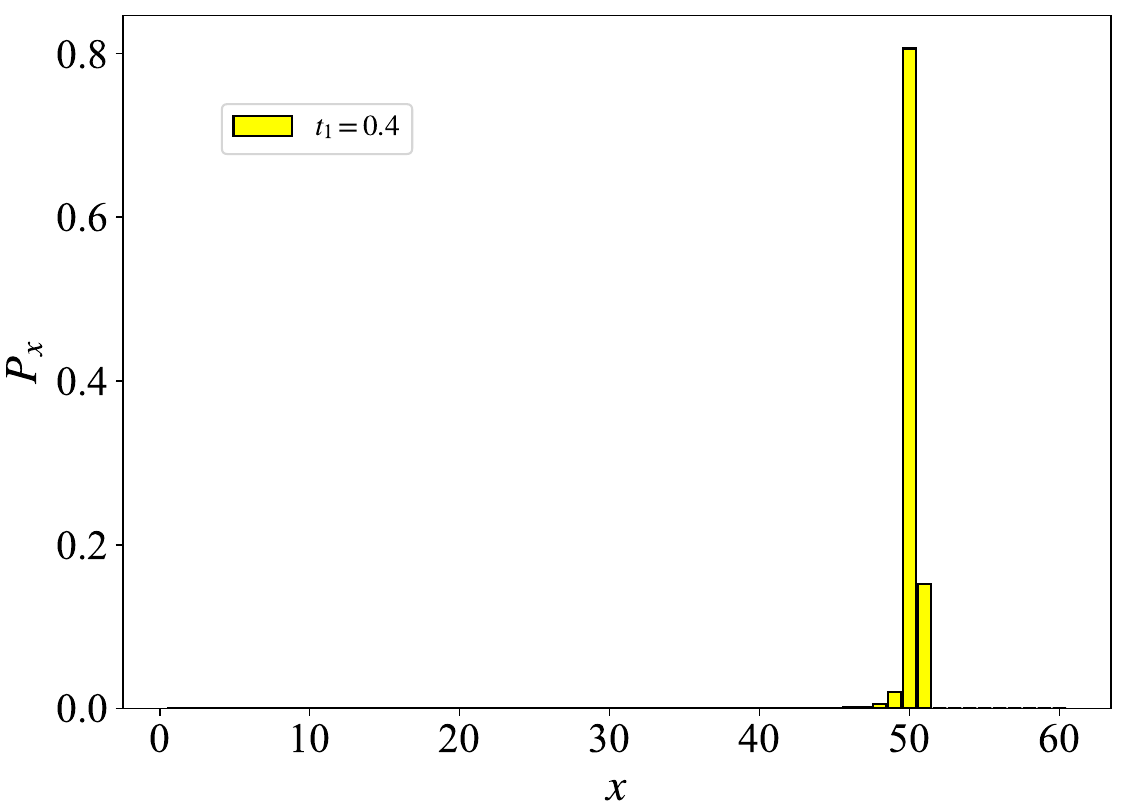}
    \caption{The spatially resolved loss probability $P_{x}$ for a walker initialized at sublattice B of site 50. 
    Other parameters are the same as those in Fig.~\ref{fig2} (a).}
    \label{fig_app2}
\end{figure} 

In the main part of the paper, we only discussed the case where the walker was initially put at some sublattice A. 
We note that such initial condition is crucial for the formation of the edge burst. For the case where the walker starts 
from some sublattice B, the edge burst is missing even though the system features NHSE and the imaginary gap closing condition.

As a simple example, we showed this phenomenon in Fig.~\ref{fig_app2}, where 
the system shares the same parameters as those in Fig.~\ref{fig2} (a) 
except that the walker is initialized at sublattice B of site 50. 
The dissipation probability of site $B_{1}$ is negligible small in this case due to the fast 
dissipation process at the initial site $B_{50}$, which can be verified by our time-perturbation analysis. 
Such result is consistent with the fact that the solution of the dynamical equation (\ref{eq1}) depends heavily on the initial condition.

\section{wave functions of the quantum walk with a non-uniform loss rate}\label{Appendix D}
\begin{figure}[tbp]
    \centering
    \includegraphics[width=0.35\textwidth]{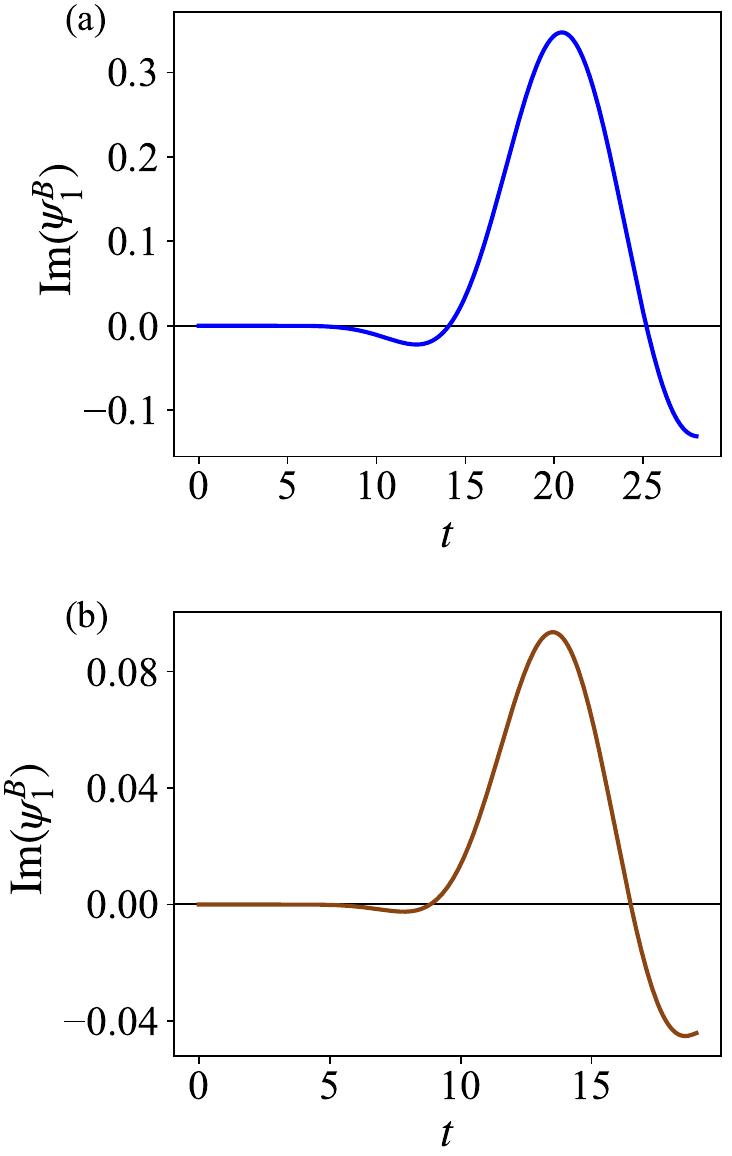}
    \caption{(a) and (b) are the edge wave functions $\psi_{1}^{B}\left(t\right)$ with perturbation calculation up to 56-th order. 
    The parameters are ${\gamma _x} = 0.2x, {t_1} = 0.4, {t_2} = 0.5$ for (a) and ${\gamma _x} = 0.2x, {t_1} = 0.8, {t_2} = 0.5$.}
    \label{fig7}
\end{figure} 

In Appendix D, we give details of calculation about the wave functions of the quantum walk 
   with a non-uniform loss rate via time-dependent perturbation theory. For simplicity, the non-uniform loss rate takes the linear form 
   ${\gamma _x} = x \gamma_0 $. Similar to the uniform loss rate case, we can apply our analytical method to this different class of problems.  
   By choosing the onsite potential operator as the unperturbed Hamiltonian ${\hat H_0}$,  
   the matrix elements under the real-space representation are 
   $\left\langle x,s\right\vert \hat{H}_{0}\left\vert x^{\prime
   },s^{\prime}\right\rangle =-\frac{i x \gamma}{2}\delta_{xx^{\prime}}%
   \delta_{ss^{\prime}}\left[  1-\left(  \sigma_{z}\right)  _{ss^{\prime}%
   }\right]  ,$ with $x,x^{\prime}$
   referring to the location of the unit cell and $s,s^{\prime}=A,B$ referring to the sublattice label. The
   hopping of the walker is treated as the perturbation operator $\hat{H}^{\prime}$ and
   its matrix elements are $H_{xs,x^{\prime}s^{\prime}}^{\prime}=\left\langle
   x,s\left\vert \hat{H}^{\prime}\right\vert x^{\prime},s^{\prime}\right\rangle
   $. Then, we can use the perturbation procedure introduced in Sec. III and evaluate the real-space wave functions. 
   For example, we evaluate the case where the system size L=12 with the walker initially prepared on 
   sublattice A of 8th site, as shown in Fig.~\ref{fig7}. The analytical results are in agreement with numerical ones.

The above example shows that we can apply the non-Hermitian time-dependent theory to 
   the systems without discrete translational symmetry.

\end{appendix}

\bibliography{mybib}

\end{document}